\documentclass[10pt,twocolumn,letterpaper]{article}

\usepackage{iccv}      
\usepackage{array}
\usepackage{tabularx}
\newcolumntype{Y}{>{\centering\arraybackslash}X}

\definecolor{iccvblue}{rgb}{0.21,0.49,0.74}
\usepackage[pagebackref,breaklinks,colorlinks,allcolors=iccvblue]{hyperref}

\def\paperID{14180} 
\def\confName{ICCV}
\def\confYear{2025}

\title{SCI-Reason: A Dataset with Chain-of-Thought Rationales for Complex Multimodal Reasoning in Academic Areas}

\author{
Chenghao Ma$^{1}$, Haihong E$^{1}$, Junpeng Ding$^{1}$, Jun Zhang$^{1}$, \\
Ziyan Ma$^{1}$, Huang Qing$^{1}$, Bofei Gao$^{2}$, Liang Chen$^{2}$, Yifan Zhu$^{1}$, Meina Song$^{1}$ \\
\\
$^{1}$Beijing University of Posts and Telecommunications \quad $^{2}$Peking University
}

\begin{document}
\maketitle
\begin{abstract}

Large Language Models (LLMs) and Large Multimodal Models (LMMs) demonstrate impressive problem-solving skills in many tasks and domains. However, their ability to reason with complex images in academic domains has not been systematically investigated. To bridge this gap, we present 
\textbf{SCI-Reason}, a dataset for complex multimodel reasoning in academic areas. SCI-Reason aims to test and improve the reasoning ability of large multimodal models using real complex images in academic domains. The dataset contains 12,066 images and 12,626 question-answer pairs extracted from PubMed, divided into training, validation and test splits. Each question-answer pair also contains an accurate and efficient inference chain as a guide to improving the inference properties of the dataset. With SCI-Reason, we performed a comprehensive evaluation of 8 well-known models. The best performing model, Claude-3.7-Sonnet, only achieved an accuracy of 55.19\%. Error analysis shows that more than half of the model failures are due to breakdowns in multi-step inference chains rather than errors in primary visual feature extraction. This finding underscores the inherent limitations in reasoning capabilities exhibited by current multimodal models when processing complex image analysis tasks within authentic academic contexts. Experiments on open-source models show that SCI-Reason not only enhances reasoning ability but also demonstrates cross-domain generalization in VQA tasks. We also explore future applications of model inference capabilities in this domain, highlighting its potential for future research.
\end{abstract}    
\section{Introduction}
\label{sec:intro}

\begin{figure}
    \centering
    \includegraphics[width=1\linewidth]{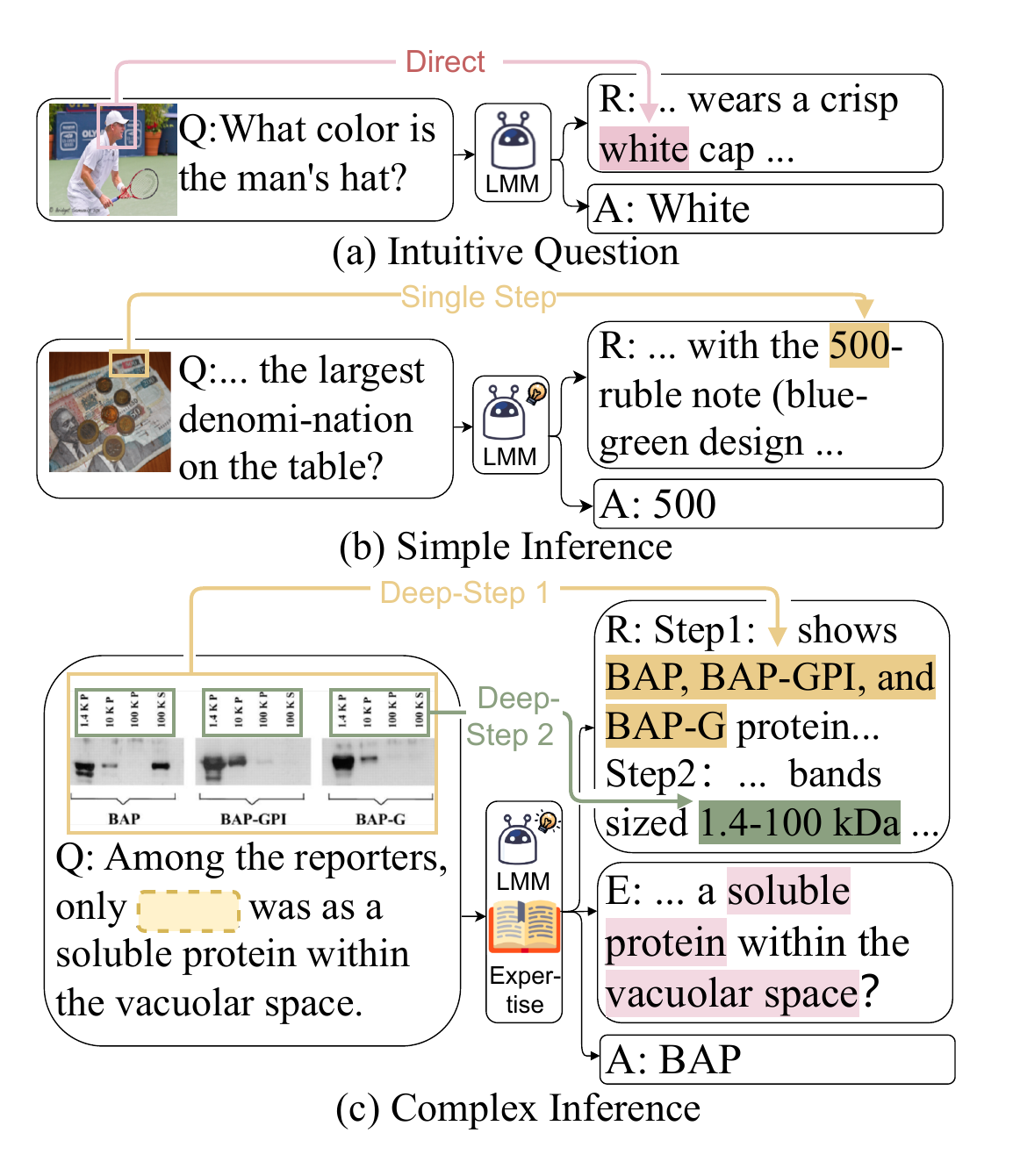}
    \caption{Comparison of conventional reasoning tasks and academic complex multimodel reasoning tasks}
    \label{fig:task_description}
\end{figure}

Recent advances in Large Multimodal Models (LMMs)\cite{bard2023,openai2023gpt4v,gemini2023} demonstrate emergent mathematical reasoning\cite{lu2022survey,kim2023athena} capabilities through techniques like chain-of-thought(CoT)\cite{wei2022chain,sprague2024cot} prompting and sparse-reward reinforcement learning\cite{guo2025deepseek,cao2024survey,team2025kimi}. These breakthroughs predominantly optimize symbolic manipulation in controlled environments, yet real-world image reasoning presents greater complexity. Complex academic multimodal reasoning involves not only numerical computation, but also cross-modal alignment of domain-specific semantics and structured visual logic parsing, as illustrated in \cref{fig:task_description}. Taking image-text consistency detection\cite{qi2024sniffer,li2024adaptive,li2024improving} in academic papers as an example, the model needs to simultaneously parse complex images containing multiple subfigures in professional academic papers and combine the information in each subfigure to check whether it supports the textual conclusions, relying on multilevel reasoning\cite{wang2024exploring,lu2022learn,yang2023mm} from pixel perception to knowledge-based hypothesis testing, which is far more complex than the single-dimensional challenges posed by current mathematical reasoning tasks\cite{ahn2024large,shao2024deepseekmath}.

Despite the growing research on multimodal reasoning datasets, existing works still exhibit three critical limitations, (1)\textbf{Domain Specialization Bias}: current datasets such as MathVista\cite{lu2023mathvista} and ChartQA\cite{masry2022chartqa} exhibit an over-reliance on symbolic mathematical reasoning (e.g., algebraic computations, geometric proofs), while neglecting broader academic scenarios that demand cross-modal knowledge integration. This narrow focus limits their ability to evaluate models in contexts requiring the synthesis of visual, textual, and symbolic information, such as interpreting biological imagery alongside chemical formulas or correlating experimental data with theoretical frameworks. (2)\textbf{Superficial Chain-of-Thought Validation}: while datasets like ScienceQA\cite{lu2022learn} and SlideVQA\cite{tanaka2023slidevqa} extend into academic applications, their evaluation frameworks lack robust mechanisms for validating reasoning processes. These benchmarks rely on linear question-answering paradigms, enabling models to exploit superficial patterns (e.g., keyword matching\cite{chen2024unified}) rather than demonstrating genuine reasoning. Furthermore, their "explanatory" annotations often lack domain-specific verification, leading to unreliable assessments of model logic and causality. (3)\textbf{Single-Image Restriction}: even advanced benchmarks like PCA-Bench\cite{chen2024pca} and MMEvalPro\cite{huang2024mmevalpro}, which incorporate reasoning-chain annotations, remain confined to single-image input paradigms. This design fails to address the multi-document analysis prevalent in real academic research, such as cross-referencing experimental figures, validating hypotheses against control-group visuals, or synthesizing complementary evidence from multiple modalities.This critical gap motivates the development of a new multimodal reasoning dataset that rigorously aligns with authentic academic contexts, integrates certified reasoning chains for verifiable logic, and transcends single-image limitations to address multi-document complexity.

To address the above issues, this paper proposes Sci-Reason, the first multimodal reasoning dataset with Chain-of-Thought rationales for complex multimodel reasoning in academic domains, whose innovations are reflected in the fine-grained knowledge-guided QA construction paradigm and the Validable Thought Chain annotation framework. Specific contributions are as follows:
\begin{itemize}
    \item We extract \textbf{high-confidence image-text pairs} from academic PubMed databases and generate \textbf{high-quality thought chain collections} using MCTS to construct SCI-Reason, a dataset featuring structured knowledge blanks for complex multimodel reasoning in academic areas.
    \item We formalize the scientific image reasoning workflow through five core competencies spanning:  \textbf{professional entity location reasoning}, \textbf{multimodal temporal reasoning}, \textbf{cross-subgraph role reasoning}, \textbf{causal mechanism reasoning}, and \textbf{methodological technical reasoning}. This taxonomy systematically encapsulates the essential challenges in complex multimodel reasoning tasks in academic areas.
    \item We tested both the test and training sets of our dataset separately, and verified that our dataset is not only able to effectively \textbf{assess the model's ability in complex multimodel reasoning tasks}, but also helpful in \textbf{improving the model's reasoning ability} in academic scenarios.
\end{itemize}

To systematically evaluate the practical value of SCI-Reason, we perform a multi-level experimental validation: first, we fine-tune the mainstream multimodal macromodel, Qwen2-VL-7B\cite{wang2024qwen2} based on the SCI-Reason training set. We also evaluate their ability to generalise across cross-domain public datasets. Moreover, we systematically assess 8 advanced models on the SCI-Reason test set, and collate the reasons for each model's errors on the test set, generalising these insights to inform future research.
\section{The SCI-Reason Dataset}
\label{sec:The SCI Dataset}

\subsection{Collection Guidelines}

\begin{figure*}
    \centering    \includegraphics[width=1\linewidth]{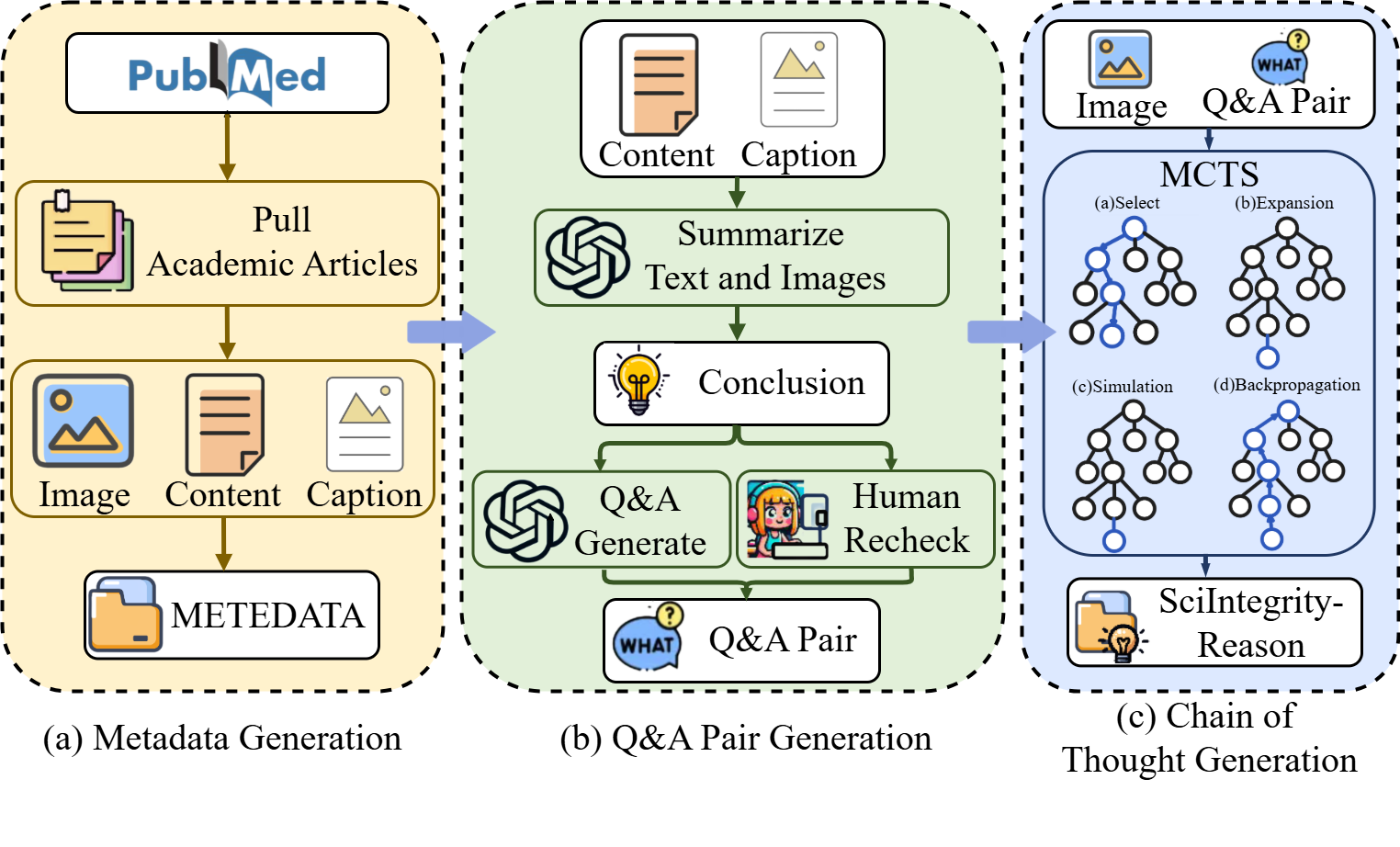}
    \caption{Overview of the dataset construction process, figure (a) illustrates the metadata collection process, figure (b) depicts the construction of question-answer pairs, figure (c) shows the generation of chain-of-thought annotations}
    \label{fig:dataset_construction}
\end{figure*}

Current multimodal models exhibit critical limitations in complex academic image reasoning, while existing benchmarks inadequately assess domain-specific analytical capabilities. To bridge this dual challenge, we develop SCI-Reason to extend multimodal reasoning beyond mathematical contexts into authentic academic applications. Our dataset adheres to four key principles: (1) \textbf{Source Authenticity}: all images and QA pairs come from peer-reviewed publications, with biomedical accuracy validated by domain experts.(2) \textbf{Structural Complexity}: multipanel composition (3-7 interconnected subfigures per image) requires cross-panel relational reasoning. (3) \textbf{Multimodal Diversity}: 4 types of academic visualization, including statistical charts, microscopic images, conceptual schematics and others, ensure real-world applicability. (4) \textbf{Efficient Chain of Thought}: an accurate and efficient chain of thought can provide guidance on model training and error analysis.

The creation of logically rigorous reasoning chains remains a fundamental challenge in complex image understanding tasks. Traditional thought chain generation approaches like generating chains of thought with stronger models\cite{chen2024huatuogpt}, often produce rationales lacking verifiable logical coherence and balanced difficulty distribution. To overcome these limitations, we implement Monte Carlo Tree Search (MCTS)\cite{coulom2006efficient} through its unique capacity to navigate combinatorial reasoning spaces via simulation-based exploration, ensuring the generation of logically consistent chains while providing stepwise granular verification signals. The algorithm's strategic tree expansion mechanism enables dynamic difficulty-aware sampling, effectively addressing the inherent bias in conventional methods toward simpler reasoning patterns.

Crucially, during data curation, MCTS's backward-chaining verification automatically detects and eliminates textual claims exceeding image-evidence support through iterative hypothesis pruning. This process establishes tight visual-textual coupling in the final reasoning chains, effectively mitigating information asymmetry between multimodal inputs. The resulting high-fidelity chain annotations provide reliable supervision signals that not only enhance model training but also enable comprehensive capability assessment through interpretable reasoning traces.

\subsection{Data Collection}

\textbf{Metadata Curation}. The data acquisition pipeline initiates by harvesting authentic academic image-text pairs from PubMed Central's open-access repository, as illustrated in figure(a) in \cref{fig:dataset_construction}. Through an automated text parsing engine equipped with domain-specific syntactic rules, we establish precise alignment between images and their corresponding methodological descriptions and experimental conclusions in full-text articles. To ensure visual complexity, a saliency-aware clustering algorithm filters multi-panel figures containing more then 4 semantically interconnected subfigures, systematically excluding simplistic single-view visuals. This process constructs foundational data triplets comprising original multi-panel figures, expert-validated image descriptions and evidence-based scientific claims.

\textbf{Question-Answer Pair Generation}. As presented in figure(b) in \cref{fig:dataset_construction}, the question-answer pair construction process employs a dual-stage knowledge refinement approach to ensure assessment quality. Initially, we perform semantic distillation of image descriptions and experimental conclusions using large language models (LLMs), extracting concise knowledge statements through iterative refinement. Subsequent contextual gap creation leverages the LLM's semantic role labeling capabilities to strategically mask domain-specific entities (methods, mechanisms, temporal markers), transforming declarative statements into fill-in-the-blank queries requiring multimodal reasoning. We have asked experts in the field to verify the quality of the questions and answers to ensure the validity of the quizzes. The resultant contextual gap queries are organized through discourse structure analysis into five specialized reasoning types: (1) professional entity location, (2) multimodal temporal reasoning, (3) cross-subgraph role reasoning (4) causal mechanism reasoning, and (5) methodological technical reasoning - collectively spanning the essential competencies for academic visual reasoning.

\subsection{CoT Generation}

The integration of verifiable reasoning trajectories into our benchmark serves dual objectives: (1) providing explicit guidance for deep analytical reasoning, and (2) enabling granular evaluation of inference validity. As shown in part(b) of the \cref{fig:dataset_construction}, we address this through Monte Carlo Tree Search (MCTS), which systematically constructs structured reasoning chains. Formally, given an academic visual reasoning instance comprising question $Q$ and image $I$, we seek an optimal reasoning path $R^* = \{r_1, r_2, ..., r_n\}$ that satisfies:
\begin{equation}
R^* = \mathop{\text{argmax}}\limits_{R \in \mathcal{P}} \prod_{i=1}^{|R|} \phi(r_i|r_{<i}, Q, I)
\end{equation}
where $\mathcal{P}$ denotes the space of all possible reasoning paths, and each $r_i$ denotes an atomic inference operation (visual observation or knowledge-based deduction) contributing to final answer $A$. The MCTS framework models this as a quadruple search tree $\mathcal{T} = (\mathcal{N},\mathcal{E},v,\pi)$ where $\mathcal{N}$ contains partial reasoning paths $R_p^{(k)} = \{r_1,...,r_k\}$, $\mathcal{E}$ represents state transitions between reasoning steps, $v(n)$ estimates stepwise correctness likelihood, and $\pi$ governs the exploration-exploitation balance during path selection. The reward function combines path fidelity (agreement with ground-truth reasoning) and explanatory completeness (coverage of required knowledge components).

Our MCTS implementation iteratively evolves the reasoning search tree through four rigorously defined phases:

\textbf{Selection Phase:}
The algorithm navigates from root to leaf node using a modified Upper Confidence Bound (UCB)\cite{garivier2011upper} formulation:
\begin{equation}
\text{UCB}(n) = \underbrace{\frac{v(n)}{N(n)}}_{\text{Exploitation}} + c \cdot \underbrace{\sqrt{\frac{\ln N(p(n))}{N(n)}}}_{\text{Exploration}}
\end{equation}
where $v(n)$ represents the \emph{cumulative value of node n}, $N(n)$ denotes the \emph{temporally discounted visit count}, $c \in [0.5,2.0]$ is adaptively tuned based on path depth to balance exploration-exploitation trade-offs and $p(n)$ is the parent of node n.

\textbf{Expansion Phase:} 
At leaf nodes, we deploy a vision-language model $\mathcal{M}_{\text{VLM}}$ to generate novel reasoning steps. The model receives contextual inputs:
\begin{equation}
\mathcal{I} = [Q; \mathcal{F}(I); R_p^{(k)}; T]
\end{equation} 
where $Q$ represents the question, $\mathcal{F}(\cdot)$ denotes image embedding extraction, $R_p^{(k)}$ stands for current reasoning path and $T$ is the prompt.

\textbf{Simulation Phase:}
Path quality evaluation combines:
\begin{equation}
V(R) = \sum_{i=1}^{k} w_i \cdot c(r_i)
\end{equation}
where $c(r_i)$ is the correctness score for step i, $w_i$ is the weight of step i and $k$ denotes the current path length.

\textbf{Backpropagation Phase:}
Node statistics update recursively through:
\begin{equation}
\begin{aligned}
v(n) &\leftarrow \frac{N(n) \cdot v(n) + V(R)}{N(n) + 1} \\
N(n) &\leftarrow N(n) + 1 
\end{aligned}
\end{equation}
This dual-update mechanism ensures both path fidelity and explanatory completeness are preserved throughout the search process.

\begin{figure}[h]
    \centering
    \begin{minipage}{0.5\textwidth}
    \centering
    \begin{tabular}{p{0.55\linewidth}>{\centering\arraybackslash}p{0.2\linewidth}}
        \toprule
        \multicolumn{1}{c}{\textbf{Statistic}} & \multicolumn{1}{c}{\textbf{Number}} \\
        \midrule
        Total question & 12,628  \\
        - Train set &  10005 \\
        - Val set & 1515 \\
        - Test set & 1107 \\
        - Multimodal temporal reasoning & 5,026(39.8\%) \\
        - Professional entity location reasoning & 3,271(25.9\%)  \\
        - Cross-subgraph role reasoning & 1,667(13.2\%) \\
        - Causal mechanism reasoning & 1,477(11.7\%) \\
        - Methodological technical reasoning & 1,187(9.4\%) \\
        \toprule
        Total images & 11805\\
        - Average subfigures per image & 6.01\\
        - Including statistical charts & 73.2\%\\
        - Including microscopic images & 47.0\%\\
        - Including schematic diagrams & 10.5\%\\
        - Including others & 11.5\%\\
        \toprule
        Unique number of answers & 9102\\
        - Average steps of CoT & 4.28\\
        - Numerical & 20.0\% \\
        - Jargon & 38.6\% \\
        - Descriptive text & 41.4\% \\
        \toprule
        Maximum question length & 243 \\
        Maximum answer length & 16\\
        Maximum CoT length & 408\\
        Average question length & 27.96\\
        Average answer length & 1.60\\
        Average CoT length & 259.93\\
        \bottomrule
    \end{tabular}
    \captionof{table}{Statistics of Scientific Reasoning Question Types}
    \label{tab:question-stats}
\end{minipage}
\end{figure}

\subsection{Data Analysis}

\begin{figure*}
    \centering
    \includegraphics[width=1\linewidth]{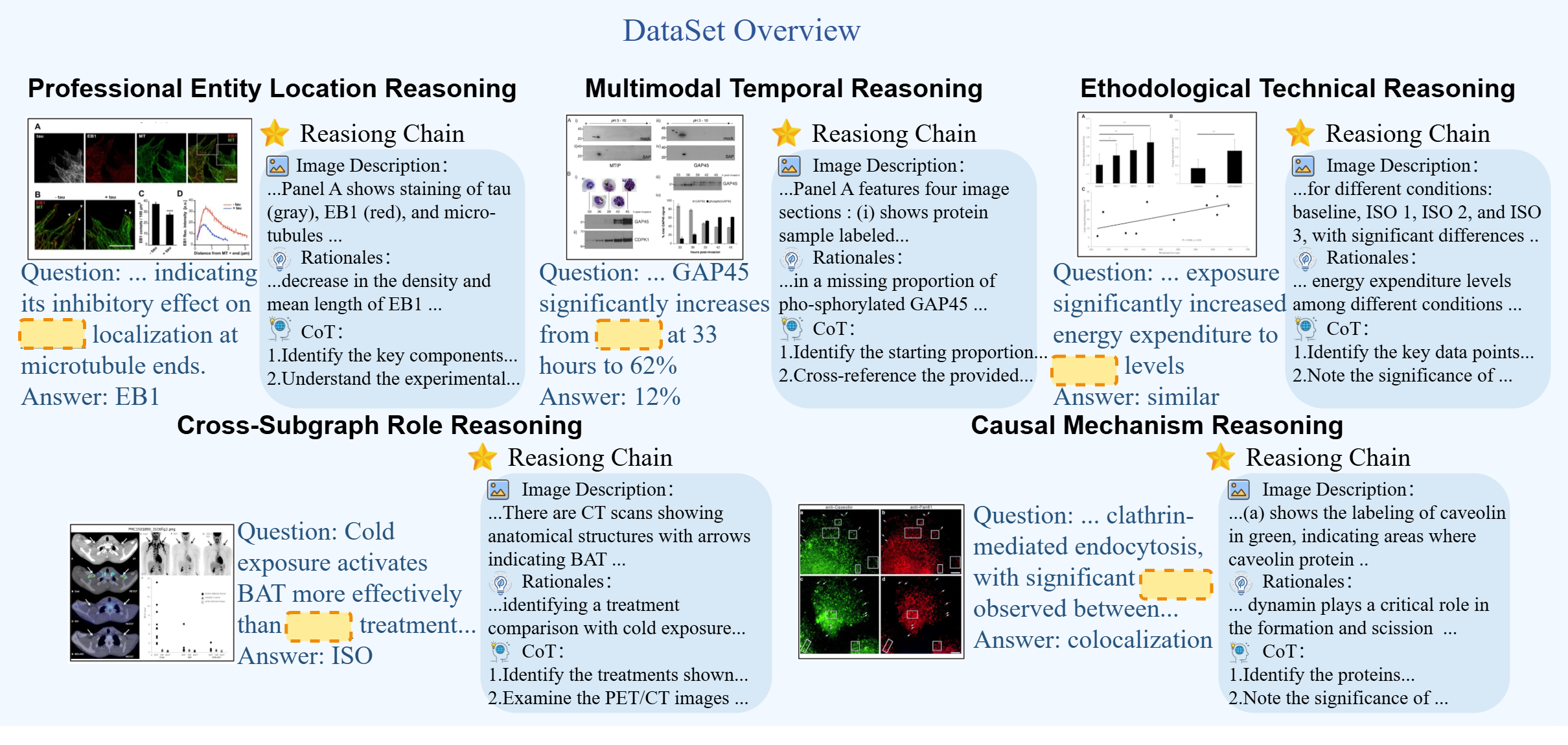}
    \caption{Examples of the five classified tasks in our dataset:multimodal temporal reasoning, professional entity location reasoning, cross-subgraph role reasoning, causal mechanism reasoning and methodological technical reasoning}
    \label{fig:key_study}
\end{figure*}

The main statistics of SCI-Reason are presented in \cref{tab:question-stats}. Given the inherent complexity and diversity of academic visual representations, our dataset analysis reveals the following composition of multi-panel figures: 73.2\% contain statistical chart subfigures (e.g., bar charts, line graphs, scatter plots), 47.0\% incorporate microscopic imaging data (including fluorescence microscopy and electron microscopy specimens), 10.5\% present conceptual schematics (such as biological pathway diagrams and mechanistic illustrations), with the remaining 11\% comprising temporal sequences and other composite visualizations. This distribution aligns with the prevalence patterns observed in STEM literature while ensuring comprehensive coverage of scientific communication modalities. This diverse distribution of image types ensures that the dataset has adequate coverage of all types of image representations in scientific papers.

In terms of the distribution of problem types, we use a five-classification system, as shown in \cref{fig:key_study}: multimodal temporal reasoning (39.8\%), professional entity location reasoning (25.9\%), cross-subgraph role reasoning (13.2\%), causal mechanism reasoning (11.7\%) and methodological technical reasoning (9.4\%). The analysis revealed clear patterns of association between different types of questions and image types. For example, the microscopic image type more often corresponded to professional entity location reasoning questions, whereas the statistical diagram type was more often involved in causal mechanism reasoning. This association reflects the characteristics of information carried by different types of images in scientific research.

As for answer characteristics, we observed a variety of answer formats, including numerical (20.0\%), jargon (38.6\%), and descriptive text (41.4\%). The average length of the answers was 1.6 words, which reflects the precision and conciseness characterising the answers in science quizzes. 
\section{Experiments}

\subsection{Evalution Protocols}

Considering the diversity of answers that may occur in open-ended fill-in-the-blank questions, we use three metrics to evaluate the effectiveness of model responses in terms of accuracy, answer similarity, and semantic similarity. Accuracy ensures correctness, answer similarity measures closeness to reference answers, and semantic similarity assesses the preservation of meaning, providing a comprehensive evaluation framework.

\textbf{Acuracy (ACC)}. First we judge the model's ability to hit the correct answer exactly by calculating the accuracy of the model's answer in hitting the correct answer. This is the most stringent evaluation criterion, which requires the model's answers to match the standard answers exactly. Although this metric is more stringent, it can intuitively reflect the model's ability to answer accurately.

\textbf{Average Normalized Levenshtein Similarity (ANLS)}. In order to avoid overly penalising the model for reasonable responses due to subtle differences in the way answers are presented, we use the ANLS\cite{peer2024anls} assessment metric. This metric provides a better measure of answer similarity by calculating the Levenshtein distance between the predicted answers and the standard answers and normalising it.The formula for ANLS is:
\begin{equation}
ANLS = 1 - \frac{LevenshteinDistance(pred, gt)}{max(len(pred), len(gt))}
\end{equation}

\textbf{WUPS (Wu-Palmer Similarity)}. Considering the importance of professional terms in scientific Q\&A, we use the WUPS (Wu-Palmer Similarity)\cite{guessoum2016modification} evaluation metric. This metric is based on the WordNet hierarchy, which can capture the semantic similarity between words, and is particularly suitable for evaluating the approximate matching of technical terms. For example, the terms ‘cell membrane’ and ‘plasma membrane’ express the same biological concept although they are literally different.

With these three complementary evaluation metrics, we are able to more comprehensively assess the performance of the model in the scientific image quizzing task. In particular, this multi-dimensional evaluation system shows its advantages when dealing with challenges specific to scientific texts such as technical terms and synonymous expressions.

\subsection{Experiment set up}

\begin{table}
  \centering
  \begin{tabular}{@{}p{35mm}p{10mm}p{10mm}p{15mm}@{}}
    \toprule
    Model & Acc & ANLS & WUPS \\
    \midrule
    Claude-3.7-Sonnet & \textbf{55.19} & \textbf{58.18} & 61.16\\
    Claude-3.7-Sonnet(with CoT) & 55.83 & 58.72 & 61.43\\
    \midrule
    GPT-4O & 50.05 & 52.30 & \textbf{61.34}\\
    GPT-4O(with CoT) & 53.66 & 57.18 & 63.32\\
    \midrule
    Gemini & 32.82 & 44.76 & 51.27\\
    Gemini(with CoT) & 49.41 & 52.94 & 55.47\\
    \midrule
    Doubao-1-5-Vision-pro & 49.50 & 54.56 & 55.01\\
    Doubao-1-5-Vision-pro(with CoT) & 49.77 & 55.65 & 56.01\\
    \midrule
    Glm-4v-Plus & 47.47 & 50.54 & 54.52\\
    Glm-4v-Plus(with CoT) & 47.15 & 49.77 & 53.93\\
    \midrule
    Step-1v-8k & 47.79 & 51.04 & 54.02\\
    Step-1v-8k(with CoT) & 47.24 & 50.95 & 53.48\\
    \midrule
    Yi-Vision-V2 & 38.12 & 39.93 & 44.72\\
    Yi-Vision-V2(with CoT) & 35.95 & 38.03 & 43.45\\
    \midrule
    Qwen-VL-Plus & 34.78 & 37.22 &41.55\\
    Qwen-VL-Plus(with CoT) & 23.13 & 25.56 &26.83\\
    \bottomrule
  \end{tabular}
  \caption{Performance of mainstream multimodal models on our test set for complex multimodel reasoning.}
  \label{tab:testresult}
\end{table}

To verify the effectiveness of the dataset for complex academic image reasoning tasks, we conducted experiments on both the training and test sets. This approach allows us to comprehensively evaluate the dataset's utility in real-world scenarios. By isolating the training set for model development and the test set for performance assessment, we ensure that our evaluation is both thorough and unbiased. This separation is crucial for understanding the dataset's strengths and limitations, and for demonstrating its practical value in advancing research in this field.

We evaluated the performance of current mainstream large-scale multimodal models, including: Claude-3.7-Sonnet, GPT-4O, Gemini-1.5-flash, Doubao-1-5-Vision-pro, Glm-4v-Plus, Qwen-VL-plus, Yi-Vision-V2 and Step-1v-8k on our test set. We employed two generation strategies: direct answer generation and CoT-enhanced generation, which required the models to produce both the final answers and the reasoning process. At the same time, we harnessed the language model to analyze the questions that the model answered incorrectly, comparing its generated CoT with the efficient reasoning chains in the dataset. This analysis helped us pinpoint and categorize the errors in the model's reasoning process into the following for points: \textbf{Knowledge-based Deficiency}, \textbf{Logical Inference Error}, \textbf{Contextual Integration Deficit} and \textbf{Visual Perception Bias}.

Besides, we fine-tune the Qwen2-VL model on our training set, compare the performance of the model before and after fine-tuning, and verify the ability of the training set to improve the multimodal large model on complex academic image reasoning tasks. In addition, we also test the ability of the model on some publicly available datasets after fine-tuning it on our dataset, thus verifying the generalisation of the dataset in enhancing the reasoning ability of multimodal large models.

\subsection{Experiment Result}

\begin{table*}
  \centering
  \begin{tabularx}{\textwidth}{@{}c*{7}{>{\centering\arraybackslash}X}@{}}
    \toprule
    \textbf{Model} & \multicolumn{3}{c}{\textbf{SCI-Reason}} & \multicolumn{4}{c}{\textbf{MMMU}} \\
    \cmidrule{2-4} \cmidrule{5-8}
    & \textbf{Acc} & \textbf{ANLS} & \textbf{WUPS}  & \textbf{Business} & \textbf{Science} & \textbf{Health} & \textbf{Tech}  \\
    \midrule
    Qwen2-VL-7B & 41.64 & 44.08 & 49.50 & 42.3 & 36.0 & 50.0 & 35.7  \\
    Qwen2-VL-7B(Lora) & 53.66 & 56.19 & 60.79 & 45.6 & 38.7 & 51.3 & 36.9  \\
    \bottomrule
  \end{tabularx}
  \caption{Performance comparison between the fine-tuned Qwen2-VL-7B model and the base model on the test set and the MMMU dataset.}
  \label{tab:ftresult}
\end{table*}

\textbf{Test Set}: We rigorously evaluated the performance of several state-of-the-art models on our comprehensive test set, which is designed to assess complex academic image reasoning capabilities. Our experimental results revealed that even the most advanced model, Claude-3.7-Sonnet, achieved a maximum acuracy score of 55.19 while GPT-4O achieved a maximum WUPS score of 61.34 on the test set, indicating substantial room for improvement in this domain. Notably, almost the performance of all models showed enhancement after incorporating thought chains, as detailed in \cref{tab:testresult}. This improvement demonstrates the effectiveness of our test set in evaluating and enhancing the reasoning abilities of models in complex academic image reasoning tasks.

However, it is important to note that some models, such as Yi, did not show significant improvement after adding CoT. This lack of enhancement can be attributed to the models' inherent limitations in understanding the images and the lack of expertise. For these models, the addition of CoT did not compensate for their fundamental shortcomings in knowledge or comprehension.

\begin{figure}
    \centering
    \includegraphics[width=1\linewidth]{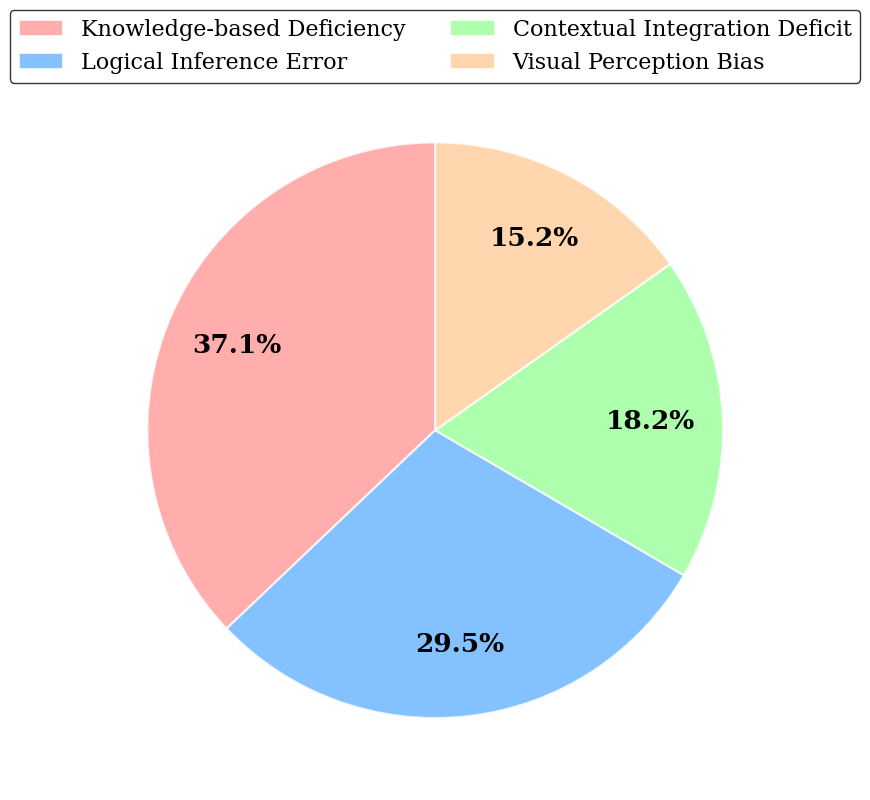}
    \caption{Analysis of model answer errors, the errors are categorized into four main types}
    \label{fig:error_analysis}
\end{figure}

Furthermore, we conducted an in-depth analysis of the thinking processes of models equipped with thought chains, as shown in \cref{fig:error_analysis}. By comparing the models' generated reasoning chains with the efficient reasoning chains in our dataset, we identified and categorized the errors in the models' reasoning processes. Error analysis shows that the main challenges of this task are reasoning rather than visual perception. Knowledge-based Deficiency(37.1\%) and Logical Inference Error (29.5\%) make up over 60\% of errors, signaling basic difficulties in scientific reasoning and knowledge use. In comparison, text and visual interpretation errors like Visual Perception Bias(15.2\%) and Contextual Integration Deficit (18.2\%) are relatively rare. This means models are decent at extracting info from figures but struggle with interpretation and logical reasoning tasks. These findings highlight the critical need for enhancing the reasoning capabilities of models in complex academic image reasoning tasks. At the same time, our analysis revealed that certain models, including Qwen-VL-Plus, exhibited significant limitations in their thinking abilities. Despite the explicit introduction of a thinking process, these models failed to develop effective thinking skills, resulting in notably inferior performance. This also highlights the inherent shortcomings in the thinking capabilities of some large multimodal models.

\textbf{Train Set}:The experimental results presented in Table~\ref{tab:ftresult} demonstrate the significant impact of fine-tuning with LoRA on the performance of the Qwen2-VL-7B model across multiple evaluation metrics and datasets. The fine-tuned model shows substantial improvements over the base model, highlighting the effectiveness of our training set in enhancing the model's reasoning abilities. The performance enhancements are evident across both the SCI-Reason test set and the MMMU dataset. On the SCI-Reason test set, the fine-tuned model achieves a notable improvement in accuracy (53.66\% vs. 41.64\%), ANLS score (56.19\% vs. 44.08\%), and WUPS score (60.79\% vs. 49.50\%) compared to the base model. These results indicate that the training set effectively equips the model with enhanced reasoning capabilities for complex academic image reasoning tasks.

Furthermore, the generalization ability of the fine-tuned model is demonstrated on the MMMU dataset. The fine-tuned model outperforms the base model across all four categories: Business (45.6\% vs. 42.3\%), Science (38.7\% vs. 36.0\%), Health (51.3\% vs. 50.0\%), and Tech (36.9\% vs. 35.7\%). This suggests that the improvements gained from training on our dataset are transferable to other related tasks and domains.

The substantial performance boost observed in the fine-tuned model underscores the unique value of SCI-Reason. SCI-Reason not only improves the model's performance on our specific task but also equips it with enhanced reasoning capabilities that are applicable to a broader range of academic image reasoning tasks. This demonstrates the practical value of our training set in advancing research in this field and highlights the importance of carefully designed training data in developing robust multimodal models.

\section{Related Work}
In recent years, Large Language Models (LLMs) have made remarkable breakthroughs in integrating symbolic reasoning with procedural knowledge. Autoregressive models represented by deepseek\cite{guo2024deepseek} and LLaMA-3\cite{dubey2024llama} have significantly improved multi-step reasoning through Chain-of-Thought (CoT) cueing and Reinforcement Fine-Tuning (RFT)\cite{zhai2025fine}. For example, deepseek improves the model's reasoning ability in mathematical reasoning by means of GRPO. SearchFormer\cite{vowinckel2023searchformer}, proposed by Tian Yundong's team, further combines symbolic search algorithms\cite{chen2023symbolic} with deep learning, demonstrating the synergistic potential of logical reasoning and path optimisation in tasks. However, most of the existing research focuses on pure text or structured symbolic reasoning (e.g., mathematical formulas, code generation), and support for multimodal scenarios is still insufficient.

Current visual question answering (VQA)\cite{antol2015vqa} benchmarks remain constrained by non-academic paradigms: Natural image datasets (e.g., A-OKVQA\cite{schwenk2022okvqa}, Visual Genome\cite{krishna2017visual}) prioritize object recognition over scholarly reasoning, while synthetic benchmarks (e.g., SuperCLEVR\cite{li2023super}, GQA\cite{hudson2019gqa}) employ oversimplified geometric primitives that lack real-world academic complexity. Recent efforts like ScienceQA\cite{lu2022learn} and MMMU\cite{yue2024mmmu} incorporate STEM diagrams but retain shallow task designs (e.g., single-image formula derivation), diverging significantly from scholarly demands for multi-modal evidence synthesis. 

Recent LLMs and LMMs have achieved remarkable results in the field of mathematical reasoning. However, this capability has not yet been widely applied to more real-world scenarios, such as the reasoning of complex academic images. The ability to reason about complex academic images is crucial, as it has significant applications in image-text consistency detection and research integrity\cite{chen2024research}.Therefore, there is an urgent need for a dataset that can apply model reasoning capabilities to a wider range of applications.
\section{Conclusion}
In this paper, we propose SCI-Reason,  a dataset for complex multimodel reasoning tasks in the academic domain, which systematically addresses the key shortcomings of existing datasets in terms of task depth, assessment reliability and domain adaptation by constructing knowledge-guided fill-in-the-blanks and Monte-Carlo Tree Search(MCTS)-driven interpretable inference chains. SCI-Reason covers a wide range of academic pictures and explicitly modelling the inference hierarchy from visual feature parsing to disciplinary logic validation. 

Our evaluation of 8 prominent foundation models highlights that while large multimodal models have achieved remarkable success in mathematical reasoning tasks, their performance in complex academic image reasoning remains limited, with the best model attaining only 55.19\% accuracy on our dataset. This underscores the need for enhanced reasoning capabilities in real-world academic scenarios. Notably, our fine-tuned Qwen2-VL-7B model surpassed GPT-4O with a 53\% accuracy on the test set, demonstrating the efficacy of SCI-Reason in improving LMM's reasoning ability. SCI-Reason offers significant potential for practical applications such as detecting inconsistent image-text content and identifying research fraud. By advancing the real-world deployment of multimodal large models, our work contributes to the development of more robust and reliable AI systems for academic and scientific research.
{
    \small
    \bibliographystyle{ieeenat_fullname}

\begin{thebibliography}{42}
\providecommand{\natexlab}[1]{#1}
\providecommand{\url}[1]{\texttt{#1}}
\expandafter\ifx\csname urlstyle\endcsname\relax
  \providecommand{\doi}[1]{doi: #1}\else
  \providecommand{\doi}{doi: \begingroup \urlstyle{rm}\Url}\fi

\bibitem[Ahn et~al.(2024)Ahn, Verma, Lou, Liu, Zhang, and Yin]{ahn2024large}
Janice Ahn, Rishu Verma, Renze Lou, Di Liu, Rui Zhang, and Wenpeng Yin.
\newblock Large language models for mathematical reasoning: Progresses and challenges.
\newblock \emph{arXiv preprint arXiv:2402.00157}, 2024.

\bibitem[Anil et~al.(2023)Anil, Borgeaud, Wu, Alayrac, Yu, Soricut, Schalkwyk, Dai, Hauth, and et~al.]{gemini2023}
Rohan Anil, Sebastian Borgeaud, Yonghui Wu, Jean-Baptiste Alayrac, Jiahui Yu, Radu Soricut, Johan Schalkwyk, Andrew~M. Dai, Anja Hauth, and et al.
\newblock Gemini: A family of highly capable multimodal models.
\newblock arXiv preprint arXiv:2312.11805, 2023.

\bibitem[Antol et~al.(2015)Antol, Agrawal, Lu, Mitchell, Batra, Zitnick, and Parikh]{antol2015vqa}
Stanislaw Antol, Aishwarya Agrawal, Jiasen Lu, Margaret Mitchell, Dhruv Batra, C~Lawrence Zitnick, and Devi Parikh.
\newblock Vqa: Visual question answering.
\newblock In \emph{Proceedings of the IEEE international conference on computer vision}, pages 2425--2433, 2015.

\bibitem[Bard(2023)]{bard2023}
Google Bard.
\newblock Google bard.
\newblock \url{https://bard.google.com/}, 2023.

\bibitem[Cao et~al.(2024)Cao, Zhao, Cheng, Shu, Chen, Liu, Liang, Zhao, Yan, and Li]{cao2024survey}
Yuji Cao, Huan Zhao, Yuheng Cheng, Ting Shu, Yue Chen, Guolong Liu, Gaoqi Liang, Junhua Zhao, Jinyue Yan, and Yun Li.
\newblock Survey on large language model-enhanced reinforcement learning: Concept, taxonomy, and methods.
\newblock \emph{IEEE Transactions on Neural Networks and Learning Systems}, 2024.

\bibitem[Chen et~al.(2024{\natexlab{a}})Chen, Cai, Ji, Wang, Liu, Wang, Hou, and Wang]{chen2024huatuogpt}
Junying Chen, Zhenyang Cai, Ke Ji, Xidong Wang, Wanlong Liu, Rongsheng Wang, Jianye Hou, and Benyou Wang.
\newblock Huatuogpt-o1, towards medical complex reasoning with llms.
\newblock \emph{arXiv preprint arXiv:2412.18925}, 2024{\natexlab{a}}.

\bibitem[Chen et~al.(2024{\natexlab{b}})Chen, Zhang, Ren, Zhao, Cai, Wang, Wang, Meng, Liu, and Chang]{chen2024pca}
Liang Chen, Yichi Zhang, Shuhuai Ren, Haozhe Zhao, Zefan Cai, Yuchi Wang, Peiyi Wang, Xiangdi Meng, Tianyu Liu, and Baobao Chang.
\newblock Pca-bench: Evaluating multimodal large language models in perception-cognition-action chain.
\newblock \emph{arXiv preprint arXiv:2402.15527}, 2024{\natexlab{b}}.

\bibitem[Chen et~al.(2023)Chen, Liang, Huang, Real, Wang, Pham, Dong, Luong, Hsieh, Lu, et~al.]{chen2023symbolic}
Xiangning Chen, Chen Liang, Da Huang, Esteban Real, Kaiyuan Wang, Hieu Pham, Xuanyi Dong, Thang Luong, Cho-Jui Hsieh, Yifeng Lu, et~al.
\newblock Symbolic discovery of optimization algorithms.
\newblock \emph{Advances in neural information processing systems}, 36:\penalty0 49205--49233, 2023.

\bibitem[Chen et~al.(2024{\natexlab{c}})Chen, Wang, Xue, Zhang, Yang, Li, Shen, Liang, Gu, and Chen]{chen2024unified}
Xiang Chen, Chenxi Wang, Yida Xue, Ningyu Zhang, Xiaoyan Yang, Qiang Li, Yue Shen, Lei Liang, Jinjie Gu, and Huajun Chen.
\newblock Unified hallucination detection for multimodal large language models.
\newblock \emph{arXiv preprint arXiv:2402.03190}, 2024{\natexlab{c}}.

\bibitem[Chen et~al.(2024{\natexlab{d}})Chen, Chen, Yang, He, Chi, Zeng, and Chen]{chen2024research}
Ziyu Chen, Changye Chen, Guozhao Yang, Xiangpeng He, Xiaoxia Chi, Zhuoying Zeng, and Xuhong Chen.
\newblock Research integrity in the era of artificial intelligence: Challenges and responses.
\newblock \emph{Medicine}, 103\penalty0 (27):\penalty0 e38811, 2024{\natexlab{d}}.

\bibitem[Coulom(2006)]{coulom2006efficient}
R{\'e}mi Coulom.
\newblock Efficient selectivity and backup operators in monte-carlo tree search.
\newblock In \emph{International conference on computers and games}, pages 72--83. Springer, 2006.

\bibitem[Dubey et~al.(2024)Dubey, Jauhri, Pandey, Kadian, Al-Dahle, Letman, Mathur, Schelten, Yang, Fan, et~al.]{dubey2024llama}
Abhimanyu Dubey, Abhinav Jauhri, Abhinav Pandey, Abhishek Kadian, Ahmad Al-Dahle, Aiesha Letman, Akhil Mathur, Alan Schelten, Amy Yang, Angela Fan, et~al.
\newblock The llama 3 herd of models.
\newblock \emph{arXiv preprint arXiv:2407.21783}, 2024.

\bibitem[Garivier and Moulines(2011)]{garivier2011upper}
Aur{\'e}lien Garivier and Eric Moulines.
\newblock On upper-confidence bound policies for switching bandit problems.
\newblock In \emph{International conference on algorithmic learning theory}, pages 174--188. Springer, 2011.

\bibitem[Guessoum et~al.(2016)Guessoum, Miraoui, and Tadj]{guessoum2016modification}
Djamel Guessoum, Moeiz Miraoui, and Chakib Tadj.
\newblock A modification of wu and palmer semantic similarity measure.
\newblock In \emph{The Tenth International Conference on Mobile Ubiquitous Computing, Systems, Services and Technologies}, pages 42--46, 2016.

\bibitem[Guo et~al.(2024)Guo, Zhu, Yang, Xie, Dong, Zhang, Chen, Bi, Wu, Li, et~al.]{guo2024deepseek}
Daya Guo, Qihao Zhu, Dejian Yang, Zhenda Xie, Kai Dong, Wentao Zhang, Guanting Chen, Xiao Bi, Yu Wu, YK Li, et~al.
\newblock Deepseek-coder: When the large language model meets programming--the rise of code intelligence.
\newblock \emph{arXiv preprint arXiv:2401.14196}, 2024.

\bibitem[Guo et~al.(2025)Guo, Yang, Zhang, Song, Zhang, Xu, Zhu, Ma, Wang, Bi, et~al.]{guo2025deepseek}
Daya Guo, Dejian Yang, Haowei Zhang, Junxiao Song, Ruoyu Zhang, Runxin Xu, Qihao Zhu, Shirong Ma, Peiyi Wang, Xiao Bi, et~al.
\newblock Deepseek-r1: Incentivizing reasoning capability in llms via reinforcement learning.
\newblock \emph{arXiv preprint arXiv:2501.12948}, 2025.

\bibitem[Huang et~al.(2024)Huang, Chen, Guo, Zeng, Zhao, Wu, Yuan, Zhao, Guo, Zhang, et~al.]{huang2024mmevalpro}
Jinsheng Huang, Liang Chen, Taian Guo, Fu Zeng, Yusheng Zhao, Bohan Wu, Ye Yuan, Haozhe Zhao, Zhihui Guo, Yichi Zhang, et~al.
\newblock Mmevalpro: Calibrating multimodal benchmarks towards trustworthy and efficient evaluation.
\newblock \emph{arXiv preprint arXiv:2407.00468}, 2024.

\bibitem[Hudson and Manning(2019)]{hudson2019gqa}
Drew~A Hudson and Christopher~D Manning.
\newblock Gqa: A new dataset for real-world visual reasoning and compositional question answering.
\newblock In \emph{Proceedings of the IEEE/CVF conference on computer vision and pattern recognition}, pages 6700--6709, 2019.

\bibitem[Kim et~al.(2023)Kim, Kim, Hahn, and Han]{kim2023athena}
JB Kim, Hazel Kim, Joonghyuk Hahn, and Yo-Sub Han.
\newblock Athena: Mathematical reasoning with thought expansion.
\newblock \emph{arXiv preprint arXiv:2311.01036}, 2023.

\bibitem[Krishna et~al.(2017)Krishna, Zhu, Groth, Johnson, Hata, Kravitz, Chen, Kalantidis, Li, Shamma, et~al.]{krishna2017visual}
Ranjay Krishna, Yuke Zhu, Oliver Groth, Justin Johnson, Kenji Hata, Joshua Kravitz, Stephanie Chen, Yannis Kalantidis, Li-Jia Li, David~A Shamma, et~al.
\newblock Visual genome: Connecting language and vision using crowdsourced dense image annotations.
\newblock \emph{International journal of computer vision}, 123:\penalty0 32--73, 2017.

\bibitem[Li et~al.(2024{\natexlab{a}})Li, Chen, Liao, and Zhang]{li2024adaptive}
Aohan Li, Jiaxin Chen, Xin Liao, and Dengyong Zhang.
\newblock Adaptive learning of consistency and inconsistency information for fake news detection.
\newblock \emph{arXiv preprint arXiv:2408.08013}, 2024{\natexlab{a}}.

\bibitem[Li et~al.(2023)Li, Wang, Stengel-Eskin, Kortylewski, Ma, Van~Durme, and Yuille]{li2023super}
Zhuowan Li, Xingrui Wang, Elias Stengel-Eskin, Adam Kortylewski, Wufei Ma, Benjamin Van~Durme, and Alan~L Yuille.
\newblock Super-clevr: A virtual benchmark to diagnose domain robustness in visual reasoning.
\newblock In \emph{Proceedings of the IEEE/CVF conference on computer vision and pattern recognition}, pages 14963--14973, 2023.

\bibitem[Li et~al.(2024{\natexlab{b}})Li, Zhang, Zhang, Zhang, and Mao]{li2024improving}
Zhe Li, Lei Zhang, Kun Zhang, Yongdong Zhang, and Zhendong Mao.
\newblock Improving image-text matching with bidirectional consistency of cross-modal alignment.
\newblock \emph{IEEE Transactions on Circuits and Systems for Video Technology}, 2024{\natexlab{b}}.

\bibitem[Lu et~al.(2022{\natexlab{a}})Lu, Mishra, Xia, Qiu, Chang, Zhu, Tafjord, Clark, and Kalyan]{lu2022learn}
Pan Lu, Swaroop Mishra, Tanglin Xia, Liang Qiu, Kai-Wei Chang, Song-Chun Zhu, Oyvind Tafjord, Peter Clark, and Ashwin Kalyan.
\newblock Learn to explain: Multimodal reasoning via thought chains for science question answering.
\newblock \emph{Advances in Neural Information Processing Systems}, 35:\penalty0 2507--2521, 2022{\natexlab{a}}.

\bibitem[Lu et~al.(2022{\natexlab{b}})Lu, Qiu, Yu, Welleck, and Chang]{lu2022survey}
Pan Lu, Liang Qiu, Wenhao Yu, Sean Welleck, and Kai-Wei Chang.
\newblock A survey of deep learning for mathematical reasoning.
\newblock \emph{arXiv preprint arXiv:2212.10535}, 2022{\natexlab{b}}.

\bibitem[Lu et~al.(2023)Lu, Bansal, Xia, Liu, Li, Hajishirzi, Cheng, Chang, Galley, and Gao]{lu2023mathvista}
Pan Lu, Hritik Bansal, Tony Xia, Jiacheng Liu, Chunyuan Li, Hannaneh Hajishirzi, Hao Cheng, Kai-Wei Chang, Michel Galley, and Jianfeng Gao.
\newblock Mathvista: Evaluating mathematical reasoning of foundation models in visual contexts.
\newblock \emph{arXiv preprint arXiv:2310.02255}, 2023.

\bibitem[Masry et~al.(2022)Masry, Long, Tan, Joty, and Hoque]{masry2022chartqa}
Ahmed Masry, Do~Xuan Long, Jia~Qing Tan, Shafiq Joty, and Enamul Hoque.
\newblock Chartqa: A benchmark for question answering about charts with visual and logical reasoning.
\newblock \emph{arXiv preprint arXiv:2203.10244}, 2022.

\bibitem[OpenAI(2023)]{openai2023gpt4v}
OpenAI.
\newblock Gpt-4v(ision) system card.
\newblock \url{https://openai.com/research/}, 2023.

\bibitem[Peer et~al.(2024)Peer, Sch{\"o}pf, Nebendahl, Rietzler, and Stabinger]{peer2024anls}
David Peer, Philemon Sch{\"o}pf, Volckmar Nebendahl, Alexander Rietzler, and Sebastian Stabinger.
\newblock Anls*--a universal document processing metric for generative large language models.
\newblock \emph{arXiv preprint arXiv:2402.03848}, 2024.

\bibitem[Qi et~al.(2024)Qi, Yan, Hsu, and Lee]{qi2024sniffer}
Peng Qi, Zehong Yan, Wynne Hsu, and Mong~Li Lee.
\newblock Sniffer: Multimodal large language model for explainable out-of-context misinformation detection.
\newblock In \emph{Proceedings of the IEEE/CVF conference on computer vision and pattern recognition}, pages 13052--13062, 2024.

\bibitem[Schwenk et~al.(2022)Schwenk, Khandelwal, Clark, Marino, and Mottaghi]{schwenk2022okvqa}
Dustin Schwenk, Apoorv Khandelwal, Christopher Clark, Kenneth Marino, and Roozbeh Mottaghi.
\newblock A-okvqa: A benchmark for visual question answering using world knowledge.
\newblock In \emph{European conference on computer vision}, pages 146--162. Springer, 2022.

\bibitem[Shao et~al.(2024)Shao, Wang, Zhu, Xu, Song, Bi, Zhang, Zhang, Li, Wu, et~al.]{shao2024deepseekmath}
Zhihong Shao, Peiyi Wang, Qihao Zhu, Runxin Xu, Junxiao Song, Xiao Bi, Haowei Zhang, Mingchuan Zhang, YK Li, Y Wu, et~al.
\newblock Deepseekmath: Pushing the limits of mathematical reasoning in open language models.
\newblock \emph{arXiv preprint arXiv:2402.03300}, 2024.

\bibitem[Sprague et~al.(2024)Sprague, Yin, Rodriguez, Jiang, Wadhwa, Singhal, Zhao, Ye, Mahowald, and Durrett]{sprague2024cot}
Zayne Sprague, Fangcong Yin, Juan~Diego Rodriguez, Dongwei Jiang, Manya Wadhwa, Prasann Singhal, Xinyu Zhao, Xi Ye, Kyle Mahowald, and Greg Durrett.
\newblock To cot or not to cot? chain-of-thought helps mainly on math and symbolic reasoning.
\newblock \emph{arXiv preprint arXiv:2409.12183}, 2024.

\bibitem[Tanaka et~al.(2023)Tanaka, Nishida, Nishida, Hasegawa, Saito, and Saito]{tanaka2023slidevqa}
Ryota Tanaka, Kyosuke Nishida, Kosuke Nishida, Taku Hasegawa, Itsumi Saito, and Kuniko Saito.
\newblock Slidevqa: A dataset for document visual question answering on multiple images.
\newblock In \emph{Proceedings of the AAAI Conference on Artificial Intelligence}, pages 13636--13645, 2023.

\bibitem[Team et~al.(2025)Team, Du, Gao, Xing, Jiang, Chen, Li, Xiao, Du, Liao, et~al.]{team2025kimi}
Kimi Team, Angang Du, Bofei Gao, Bowei Xing, Changjiu Jiang, Cheng Chen, Cheng Li, Chenjun Xiao, Chenzhuang Du, Chonghua Liao, et~al.
\newblock Kimi k1. 5: Scaling reinforcement learning with llms.
\newblock \emph{arXiv preprint arXiv:2501.12599}, 2025.

\bibitem[Vowinckel and H{\"a}hnke(2023)]{vowinckel2023searchformer}
Konrad Vowinckel and Volker~D H{\"a}hnke.
\newblock Searchformer: Semantic patent embeddings by siamese transformers for prior art search.
\newblock \emph{World Patent Information}, 73:\penalty0 102192, 2023.

\bibitem[Wang et~al.(2024{\natexlab{a}})Wang, Bai, Tan, Wang, Fan, Bai, Chen, Liu, Wang, Ge, et~al.]{wang2024qwen2}
Peng Wang, Shuai Bai, Sinan Tan, Shijie Wang, Zhihao Fan, Jinze Bai, Keqin Chen, Xuejing Liu, Jialin Wang, Wenbin Ge, et~al.
\newblock Qwen2-vl: Enhancing vision-language model's perception of the world at any resolution.
\newblock \emph{arXiv preprint arXiv:2409.12191}, 2024{\natexlab{a}}.

\bibitem[Wang et~al.(2024{\natexlab{b}})Wang, Chen, Han, Lin, Zhao, Liu, Zhai, Yuan, You, and Yang]{wang2024exploring}
Yiqi Wang, Wentao Chen, Xiaotian Han, Xudong Lin, Haiteng Zhao, Yongfei Liu, Bohan Zhai, Jianbo Yuan, Quanzeng You, and Hongxia Yang.
\newblock Exploring the reasoning abilities of multimodal large language models (mllms): A comprehensive survey on emerging trends in multimodal reasoning.
\newblock \emph{arXiv preprint arXiv:2401.06805}, 2024{\natexlab{b}}.

\bibitem[Wei et~al.(2022)Wei, Wang, Schuurmans, Bosma, Xia, Chi, Le, Zhou, et~al.]{wei2022chain}
Jason Wei, Xuezhi Wang, Dale Schuurmans, Maarten Bosma, Fei Xia, Ed Chi, Quoc~V Le, Denny Zhou, et~al.
\newblock Chain-of-thought prompting elicits reasoning in large language models.
\newblock \emph{Advances in neural information processing systems}, 35:\penalty0 24824--24837, 2022.

\bibitem[Yang et~al.(2023)Yang, Li, Wang, Lin, Azarnasab, Ahmed, Liu, Liu, Zeng, and Wang]{yang2023mm}
Zhengyuan Yang, Linjie Li, Jianfeng Wang, Kevin Lin, Ehsan Azarnasab, Faisal Ahmed, Zicheng Liu, Ce Liu, Michael Zeng, and Lijuan Wang.
\newblock Mm-react: Prompting chatgpt for multimodal reasoning and action.
\newblock \emph{arXiv preprint arXiv:2303.11381}, 2023.

\bibitem[Yue et~al.(2024)Yue, Ni, Zhang, Zheng, Liu, Zhang, Stevens, Jiang, Ren, Sun, et~al.]{yue2024mmmu}
Xiang Yue, Yuansheng Ni, Kai Zhang, Tianyu Zheng, Ruoqi Liu, Ge Zhang, Samuel Stevens, Dongfu Jiang, Weiming Ren, Yuxuan Sun, et~al.
\newblock Mmmu: A massive multi-discipline multimodal understanding and reasoning benchmark for expert agi.
\newblock In \emph{Proceedings of the IEEE/CVF Conference on Computer Vision and Pattern Recognition}, pages 9556--9567, 2024.

\bibitem[Zhai et~al.(2025)Zhai, Bai, Lin, Pan, Tong, Zhou, Suhr, Xie, LeCun, Ma, et~al.]{zhai2025fine}
Simon Zhai, Hao Bai, Zipeng Lin, Jiayi Pan, Peter Tong, Yifei Zhou, Alane Suhr, Saining Xie, Yann LeCun, Yi Ma, et~al.
\newblock Fine-tuning large vision-language models as decision-making agents via reinforcement learning.
\newblock \emph{Advances in Neural Information Processing Systems}, 37:\penalty0 110935--110971, 2025.

\end{thebibliography}

}

\end{document}